\documentclass[aps,pre,psfig,twocolumn,showpacs,superscriptaddress,reprint,floatfix]{revtex4-2}
\usepackage{amsmath}
\usepackage{mathtools}
\usepackage{amssymb}
\usepackage{bm}
\usepackage{xcolor}
\usepackage{graphics}
\usepackage{epsfig}
\usepackage[normalem]{ulem} 
\usepackage{cancel}
\usepackage[mathlines]{lineno}
\usepackage{xtab,afterpage,longtable}
\usepackage{booktabs}   
\usepackage{ltablex}
\usepackage{nicefrac}
\usepackage{float}
\usepackage{array}
\newcolumntype{C}[1]{>{\centering\let\newline\\\arraybackslash\hspace{0pt}}m{#1}}

\begin{document}
\title{Dynamically phase-separated states in driven binary dusty plasma
}
\author{Farida Batool}
\email{farida.batool@iitjammu.ac.in}
\author{Sandeep Kumar}
\author{Sanat Kumar Tiwari}
\email{sanat.tiwari@iitjammu.ac.in}
\affiliation{Department of Physics, Indian Institute of Technology Jammu, Jammu, J\&K, 181221,  India}
\date{\today}
\begin{abstract}
We comprehensively study external forcing-driven dynamical structure formation in a binary dusty plasma mixture. Using two-dimensional driven-dissipative molecular dynamics simulations, we demonstrate phase segregation into bands and lanes beyond a critical forcing threshold. The particles interact via the Debye-Hückel potential, with interaction strength serving as a control parameter for determining the critical forcing. During early evolution, the results exhibit features of two-stream instability. A steady-state phase-space diagram indicates that bands and lanes emerge beyond a critical forcing and coupling strength. Lanes predominantly form under high external forcing.
Multiple independent diagnostics, including the order parameter, drift velocity, diffusion coefficients, domain size, and the final-to-initial coupling strength ratio, provide insight into phase segregation and help determine the critical forcing amplitude. Furthermore, we show that the time evolution of band and lane widths follows an exponent of 1/3 for both critical and off-critical mixtures. These findings contrast with the previously reported scaling of 1/2 for equilibrium phase separation in critical mixtures. These results help bridge the gap between dusty plasmas and colloidal systems and facilitate controlled dusty plasma experiments in this direction.
\end{abstract}
\maketitle
\section{Introduction}
\label{intro}
\paragraph*{}
Dusty plasmas exhibit rich and complex phase separation behavior under both equilibrium and non-equilibrium conditions~\citep{Ivlev_2008_PRL, Sutterlin_2009_PRL, Wysocki_PRL_2010, Ivlev_2009_EPL,Jiang_EPL_2011,Jiang_2010_EPL,Sutterlin_2010_IOP,Du_2012_EPL,Du_2012_NJP,Killer_PRL_2016,Huang_2019_JI}.
The equilibrium scenario typically represents a first-order phase transition, where an interface separates the two phases as the system evolves. In contrast, a non-equilibrium scenario, typically driven by a continuous external force, can lead to a first-order phase transition, resulting in the formation of organized structures such as lanes or bands~\citep{Sutterlin_2009_PRL,Dzubiella_2002_PRE,Sarma_2020_PoP,Couzin_2003_PNS,Feliciani_2016_PRE}. This paper focuses on the force-driven phase separation (segregation) in a binary dusty plasma mixture. Our comprehensive study specifically points out the role of coupling strength in guiding the segregation process. The presented results unify different approaches, including the order parameter, diffusion coefficients, and the coupling strength evolution. Independently, these diagnostics predict almost the same critical forcing required to instigate band and lane formation. We also provide domain size scaling for critical and off-critical mixtures that come out as 1/3, in contrast to the equilibrium phase separation scaling of 1/2.
\paragraph*{}
In dusty plasma experiments, phase separation has been explained either through spinodal decomposition, imbalance in plasma forces, or through external driving~\citep{Wysocki_PRL_2010,Ivlev_2009_EPL,Jiang_EPL_2011,Killer_PRL_2016,Huang_2019_JI,Stephan_PRE_2020}. In some cases, the formation of the lane acts as the onset of equilibrium phase separation~\citep{Sutterlin_2009_PRL,Ivlev_2009_EPL,Du_2012_EPL,Du_2012_NJP}.
We divide the past studies of dust particles' alignment (laning) into two scenarios: whether the dust cloud is made of one type of particle~\citep{Schweigert_1996_PRE, Kroll_2010_PoP, Takahashi_1998_PRE, Arp_2012_PRE, Mendoza_2024_AXV, Joshi_2023_PRE} or is a mix of two types (binary)~\citep{Sutterlin_2009_PRL, Jiang_2010_EPL, Du_2012_EPL, Du_2012_NJP}. For both scenarios, the alignment of dust particles has been attributed to effects such as the ion-wake effect, external perturbations like an electric field~\cite{Ivlev_2008_PRL}, and the interaction between two counter-propagating dust particles. For single-type particles, Arp \textit{et al.} demonstrated chain-like dust structures formed under microgravity conditions~\citep{Arp_2012_PRE}. The explanation lies in the anisotropic interactions between the dust particles due to ion flow in its direction. Later,~\citet{Joshi_2023_PRE} included the wake effect in the interaction potential and demonstrated the dust string-like structure in simulations. 
\citet{Sutterlin_2009_PRL} demonstrated laning formation in a binary dust mixture when smaller particles cloud penetrate the background of heavier particles due to size-dependent plasma force. In a longer timescale, both types are phase-separated. Jiang \textit{et al.} investigated the effect of nonadditive interactions and initial spatial configurations on lane formation in a system with two different dust particles. They studied this behavior through simulations~\citep{Jiang_2010_EPL}. In addition to dusty plasmas, lane formation in pair-ion plasmas has been studied using Langevin dynamics~\citep{Sarma_2020_PoP, Baruah_2021_JPP}.
The present work investigates lane formation in a binary dusty plasma driven by an external force. It explores how varying plasma parameters, strong coupling, and forcing strengths influence these lanes' emergence and dynamics. 
\paragraph*{}
 We explore whether segregation can occur in a (strongly coupled) dusty plasma mixture under homogenous background when subjected to an external perturbation. Specifically, we will focus on inducing segregation through the application of laser pressure or electric fields to better understand the mechanisms underlying this phenomenon. Laser radiation pressure has been effectively used to generate shear flow in 2D dusty plasma experiments. For instance,~\citet{Feng_2012_PRL,Feng_2012_PRE} demonstrated the creation of a stable subsonic laminar flow using two counter-propagating laser beams. In comparison, our study calculates the laser pressure acting on two dust particles to achieve relative velocities, facilitating the formation of lane or band structures. This will be further elaborated in the discussion section~\ref{sec:discus}. Likewise, the application of the external electric field creates a string-like structure in an electrorheological complex plasma~\citep{Ivlev_2008_PRL}.
 Here are a few scenarios to achieve or induce phase separation in a dusty plasma experiment. For example, it is possible in a three-dimensional polydispersive dusty plasma cloud with particles of different sizes (different masses). These particles will then have a charge distribution based on their sizes~\citep{Sutterlin_2009_PRL}. If we apply an external electric field, it shall drive particles of different charges with different forces. Soon, they will acquire different terminal velocities depending on the forcing amplitude. This may lead to the lane and band formation. Laser irradiation can achieve the same as different radiation pressures, adding different velocities to multiple-sized particles. Earlier dusty plasma experiments of binary mixture achieved this by preparing a dust cloud of heavy particles and injecting the beam of light particles into the cloud~\citep{Sutterlin_2009_PRL, Du_2012_EPL, Du_2012_NJP}.
Alternatively, same-sized particles streamed through the dust cloud, forming a lane~\citep{Mendoza_2024_AXV,Arp_2012_PRE}.
\paragraph*{}
The lane is a one-dimensional array of dust particles of the same species in the binary mixture. These lanes form in the direction of forcing. Their length can range from a few particles up to the size of a dust cloud or the system size in the simulation. Further, a band is characterized as a two-dimensional structure. It is typically a stack of many lanes with a thickness of several interatomic separations. In the literature on colloids, the formation of bands can take any direction, irrespective of the forcing direction~\citep{Wysocki_2009_PRE}. The present work discusses the band only in the direction of the forcing.
\paragraph*{}
Lane formation characterised by unidirectional flow and anisotropic structural order is a non-equilibrium phenomenon. It is observed in many systems like colloids, granular media, and plasmas~\citep{Sutterlin_2009_PRL,Dzubiella_2002_PRE,Ciamarra_2005_IOP,Sarma_2020_PoP}. We can also observe lane formation in nature, such as in ant colonies and pedestrian traffic~\citep{Couzin_2003_PNS,Feliciani_2016_PRE}. Inspired by experimental studies of lane formation in dusty plasmas under microgravity, we conducted an in-depth study of Langevin molecular dynamics of this process in strongly coupled dusty plasma (SCDP). We also wish to get a better insight into phase separation dynamics beyond our previous work on phase separation in binary mixtures of strongly coupled plasma, where inhomogeneity in background screening caused phase separation~\citep{Farida_2024_PoP}.
\paragraph*{}
While studying the problem of lane formation via segregation due to external force, we also revisit the potential connection of this problem with two-stream instability (TSI). We realised that during very early evolution stages, while the streaming velocities of the species are close to satisfying the criteria of the relative velocity with other species, we may obtain a signature of the TSI, which we will discuss in the pre-segregation stage.
\paragraph*{}
This paper is organized as follows. Sec.~\ref{sec:phy_sys} outlines the physical model, including simulation parameters. Section~\ref{sec:observe} presents the observations and analysis. The analysis includes calculating the effect of the external force on the coupling strength, the order parameter, the diffusion coefficient, and the scaling of domain growth in the mixture. Additionally, Section~\ref{sec:observe} discusses the effect of coupling strength and the concentration effect on drift velocity. Finally, Sections~\ref{sec:discus} and~\ref{sec:summary} provide the discussion and summary, respectively.
\section{Model description and simulation parameters}
\label{sec:phy_sys}
\paragraph*{}
The binary mixture consists of two types of dust particles assigned as types A and B, with a mass ratio of $m_B/m_A = 2$, where $m_A$ and $m_B$ are the masses for types A and B, respectively. To obtain a two-dimensional dust layer, we set $Z_B = (m_B/m_A) Z_A$, where $Z_A$ and $Z_B$ are integer multiples of the elementary charge for the two dust species. The particles interact through screened Coulomb interaction with a uniform background screening, which is given by: 
\begin{equation}
   U_{ij} = \frac{1}{4 \pi \epsilon_0} \frac{Q_iQ_j}{r_{ij}} \exp{\left(-r_{ij}/\lambda_D \right)}.
    \label{potential_eq}
\end{equation}
In the above expression, $Q_{i,j}$ is the charge on the species, the index $i$ and $j$ are used to specify type $A$ and $B$ particles, respectively, and $\lambda_D$ is the plasma Debye length. The exponential factor quantifies the shielding of the dust particles by the background electrons and ions.
\begin{figure}[h!]
\includegraphics[width=\linewidth]{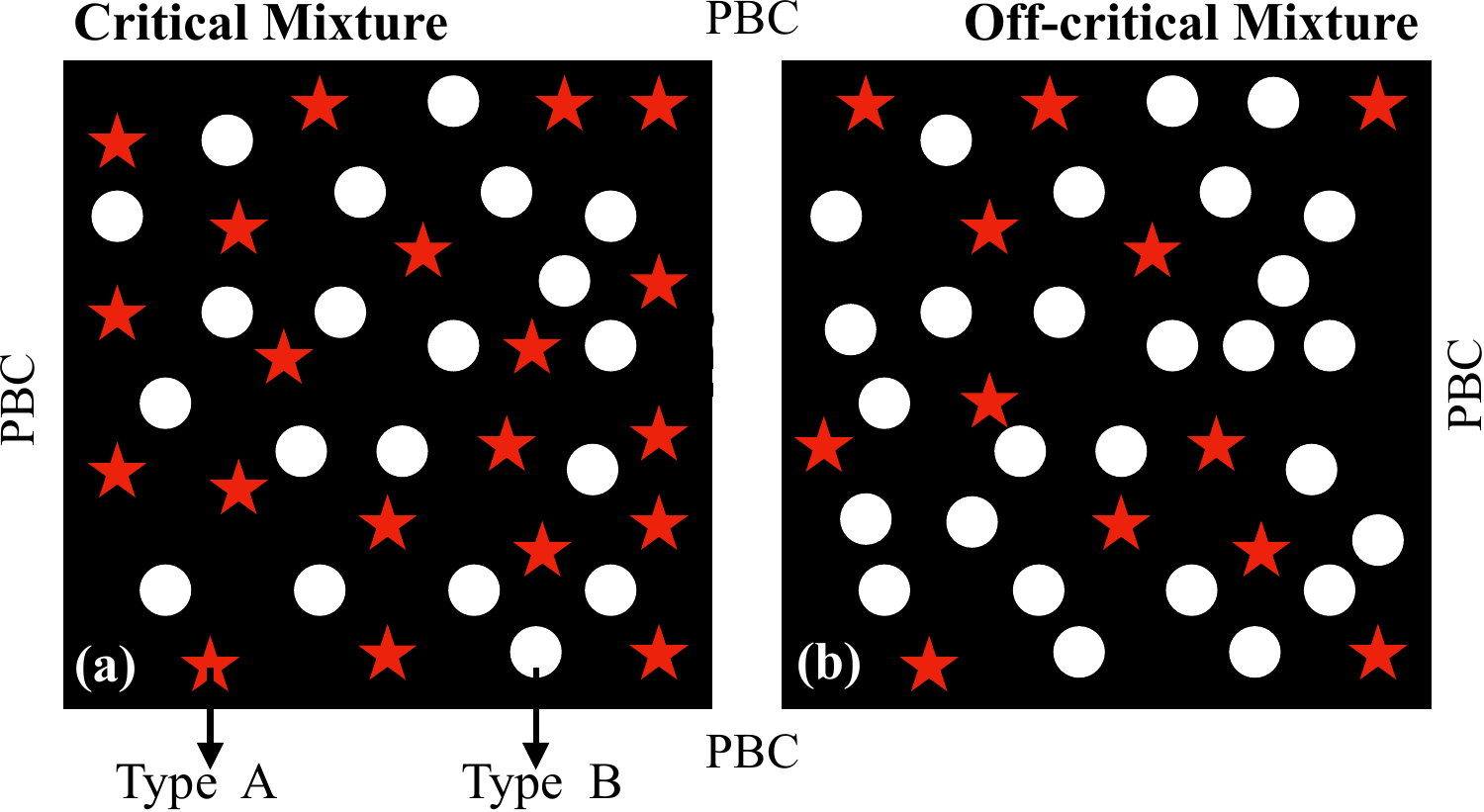}
\caption{The schematic of the simulation setup. Subplot (a) illustrates box with an equal concentration of A and B species with $\Psi_A:\Psi_B = 0.5:0.5$ and subplot (b) illustrates box with an unequal concentration of A and B species with $\Psi_A:\Psi_B = 0.3:0.7$, where $\Psi_A$ and $\Psi_B$ is $N_A/N$ and $N_B/N$ respectively.}
\label{fig:Fig1}
\end{figure}
\paragraph*{}
 We use the open-source large-scale atomic/molecular massively parallel simulator (LAMMPS) software package to conduct the simulation study~\citep{Thompson_2022_CPC}. Molecular dynamic simulation allows us to conduct the simulation at the atomistic level, making it possible to study collective behavior at different length scales. Further, it is a realistic model to study the strongly coupled plasmas (SCPs); the alternative hydrodynamic models used to study SCPs have limited applicability.~\citep{Donko_2019_AJP}. Table~\ref{tab:table1} lists the physical parameters used in the simulations, which are chosen to closely match the properties of dust particles employed in experiments~\cite{Bockwoldt_2014_PoP}. Figure~\ref{fig:Fig1} shows cartoons of a typical simulation system chosen for present studies with periodic boundaries at all sides. Subplots (a) and (b) show the binary mixture of two types of dust particles with equal (critical) and unequal (off-critical) concentrations, respectively.
\begin{table}[ht!]
{\renewcommand{\arraystretch}{1.5}
\begin{center}
\caption{The parameters used for the simulation.}
\begin{tabular}{l l}
\hline
\hline
Simulation parameters & Value (SI units) \\
\hline
$m_A$ (Mass of Type A particles)  & $6.9 \times 10^{-13}$ Kg \\
$m_B$ (Mass of Type B particles)  & $1.38 \times 10^{-12}$ Kg \\
$Q_A = Z_A e^-$ (Charge on Type A) & $12000e^{-}$ \\
$Q_B = Z_B e^-$ (Charge on Type B) & $24000e^{-}$ \\
$\kappa$ (Background screening parameter) & 0.9\\
$\omega_{pA}$ (Plasma frequency) & 51.8 s$^{-1}$ \\
$\gamma$ (Drag coefficient) & 111.1 s$^{-1}$ \\
$n$ (Total density = $n_A + n_B$) & $2.9 \times 10^6$ m$^{-2}$ \\
$a = \sqrt{(1/\pi n)}$ (Inter-particle separation) & $3.3 \times 10^{-4}$ m \\
Length of square system $(L=L_X=L_y)$ & 0.052 m ($\approx$ 158$a$)\\
$N = N_A + N_B$ & 8000 \\
\hline
\hline
\end{tabular}
\label{tab:table1}
\end{center}}
\end{table}
\paragraph*{}
An important property of the dusty plasmas is their ability to show traits of different phases. It is due to the possibility of variation in the interaction energy compared to the thermal energy of particles. This ability is quantified by the coupling parameter $\Gamma = <Q^2/a>/<k_BT>$, the ratio of average Coulomb potential energy to the average thermal energy of dust particles. As we are considering two different types of particles in the binary mixture, we have three different values of coupling strengths involved as $ \Gamma_{AA},~\Gamma_{BB}$, and  $\Gamma_{AB}$. In this paper, we only define $\Gamma_A = \Gamma_{AA}$ for mixture, and other values can be extracted as we have already set $Z_B = (m_B/m_A) Z_A$ .
In the simulations, the value of the strong coupling parameter $\Gamma$ of the particles is increased by decreasing the temperature.
\paragraph*{}
The system is equilibrated at a given temperature before the application of the external forcing. 
During the equilibration, the total force in the system consists of Debye-H\"uckel and dissipative forces:
\begin{equation}
    F = F_{DH} + F_{diss}+ F_{rand}, 
    \label{force_eqb}
\end{equation}
in which $F_{DH}$ is the pairwise interaction force between the particles defined by Eq.~\ref{potential_eq} and $F_{diss}$ is the drag force experienced by the dust particles due to interaction with the background medium. This force has a form of ‘$-\beta v$’, where $\beta$ is the damping factor equal to $m_A\gamma$. As the amplitude of the damping factor affects the formation of lanes, we reasonably chose the $\beta$ value that typically aligns with the dusty plasma experiments. In addition, the force $F_{rand}$ contributes to the random kicks experienced by the dust particles due to the background medium. Mathematically, these kicks are Gaussian random numbers with zero mean and the variance~\citep{Jiang_2010_EPL}  given by: 
 \begin{equation}
<F_{i}(t)F_{i}(t') > = 2k_BT\beta\delta(t-t').
    \label{force_rand}
\end{equation}
\paragraph*{}
After equilibration, we apply an additional external force $F_{ext}$ on type A particles to create drift. 
The external forcing $F_{ext}$ has been taken of the following mathematical form, 
\begin{equation}
    F_{ext}(t) = \begin{cases} F_{0} , & \text{if } {\text{mod}~(t, n_{dr}dt~) = 0} \\ 0, & \text{otherwise}. \end{cases} 
    \label{force_ext}
\end{equation}
Here $F_0$ is the constant forcing amplitude, $dt$ is the timestep, $t$ is the simulation time varying from 0 to the simulation duration $T_f$, and $F_{ext}$ is the external force on the type A species. In Eq.~\ref{force_ext}, $n_{dr} dt$ denotes the interval at which the force pulses are applied, and $dt$ is the simulation time step. The corresponding driving frequency $f_{dr} = 1/(n_{dr} dt)$. The present simulations have been carried out with $n_{dr} = 3$, except for Table~\ref{tab:table2}, where the value of $n_{dr}$ has been varied to study its effect on structure formation.
\section{Observation and Analysis}
\label{sec:observe}
\paragraph*{}
On the basis of the segregation stage, we broadly classify observations into two categories. 
\paragraph*{}
{\it \textbf {Pre-segregation:}} 
In this stage, particles start experiencing the external force but with no signature of lane/band formation in systems. However, during this early evolution, we observe signatures of two-stream instability transiently~\citep{Williams_2019_PRE,Chen_1984_S}. This instability arises when two charged streams interpenetrate, i.e., show relative streaming motion. In the present work, two species have different masses and relative streaming velocities, which creates conditions for TSI.
\begin{figure*}[ht!]
\includegraphics[width=\textwidth]{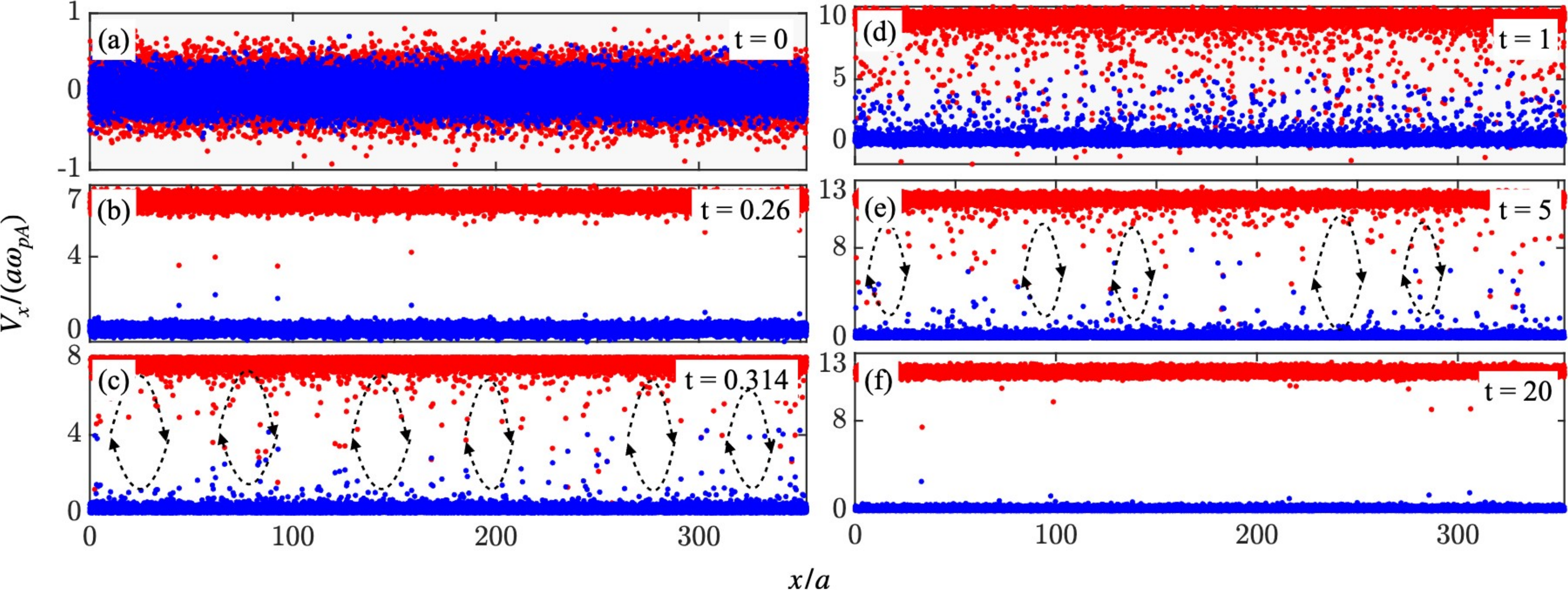}
\caption{The phase-space trajectories of the particles in the critical mixture show signatures of the TSI at early evolution. Simulation parameters are $F_{ext} = 73$ pN, $\Gamma_A = 10$, and $\kappa = 0.9$. The system size is 2.23 times the length of the system that is provided in Table~\ref{tab:table1}.}
\label{fig:Fig2}
\end{figure*}
\paragraph*{}
The phase space evolution of both types of particles [type A (red) and type B (blue)] is shown in Fig.~\ref{fig:Fig2}. During very early times, the streams almost overlap as there is little velocity difference; see Fig.~\ref {fig:Fig2} (a). The two streams are visible with exchange particles at times less than one plasma period [Fig.~\ref{fig:Fig2} (b, c)]. At plasma period durations, two stream instability-like scenarios form and can be seen in Fig.~\ref{fig:Fig2} (d, e). As the forcing continuously accelerates particles, the velocity difference between the streams increases over time, eventually becoming too large to sustain TSI in the medium. 
\paragraph*{}
{\it \textbf {Early segregation and the saturation stage:}} In this stage, the particles acquire velocity depending on the amplitude of the forcing. For a very small force, all velocity enhancement goes into the thermal velocity due to collisions; hence, we can't observe any lane/band formation. With further increase in force amplitude, the particle acquires drift velocity. Yet, there is a fraction of kinetic energy that distributes into the thermal energy due to collision; hence, velocity is not completely equivalent to the drag limit of velocity $v_d = F_{ext}/\beta$. For these values of forcing, the system starts to form small structures aligned in the direction of forcing. The collision effects become negligible at sufficiently high forcing amplitudes and the particle velocity approaches the drag limit. In this regime, clear segregation of particles along the direction of forcing is observed.
Figure~\ref{fig:fig3} shows the variation of the drift velocity $v_d$, calculated from the average particle velocity, as a function of the external force $F_{ext}$. The drift velocity is measured once the system reaches a steady state. Diamond markers represent simulation data, while the black dashed line indicates the theoretical drag limit. At low forcing values, $v_d$ remains below both the drag limit and the thermal velocity. As $F_{ext}$ increases, $v_d$ exceeds the thermal velocity and approaches the drag limit. The inset highlights the low-force regime, where the blue dashed line represents the thermal velocity $v_{th}$.
\begin{figure}[t]
\includegraphics[width=\linewidth]{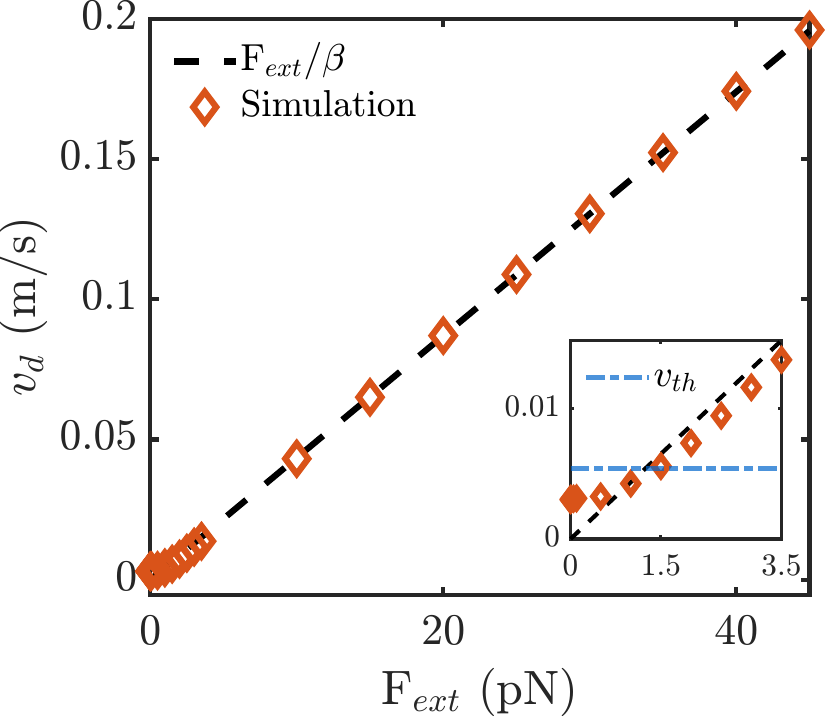}
\caption{The variation in the drift velocity with $F_{ext}$ for critical mixture. The dashed black line is the theoretical value of the drift velocity.}
\label{fig:fig3}
\end{figure}
\begin{figure}[t]
\includegraphics[width=0.5\textwidth]{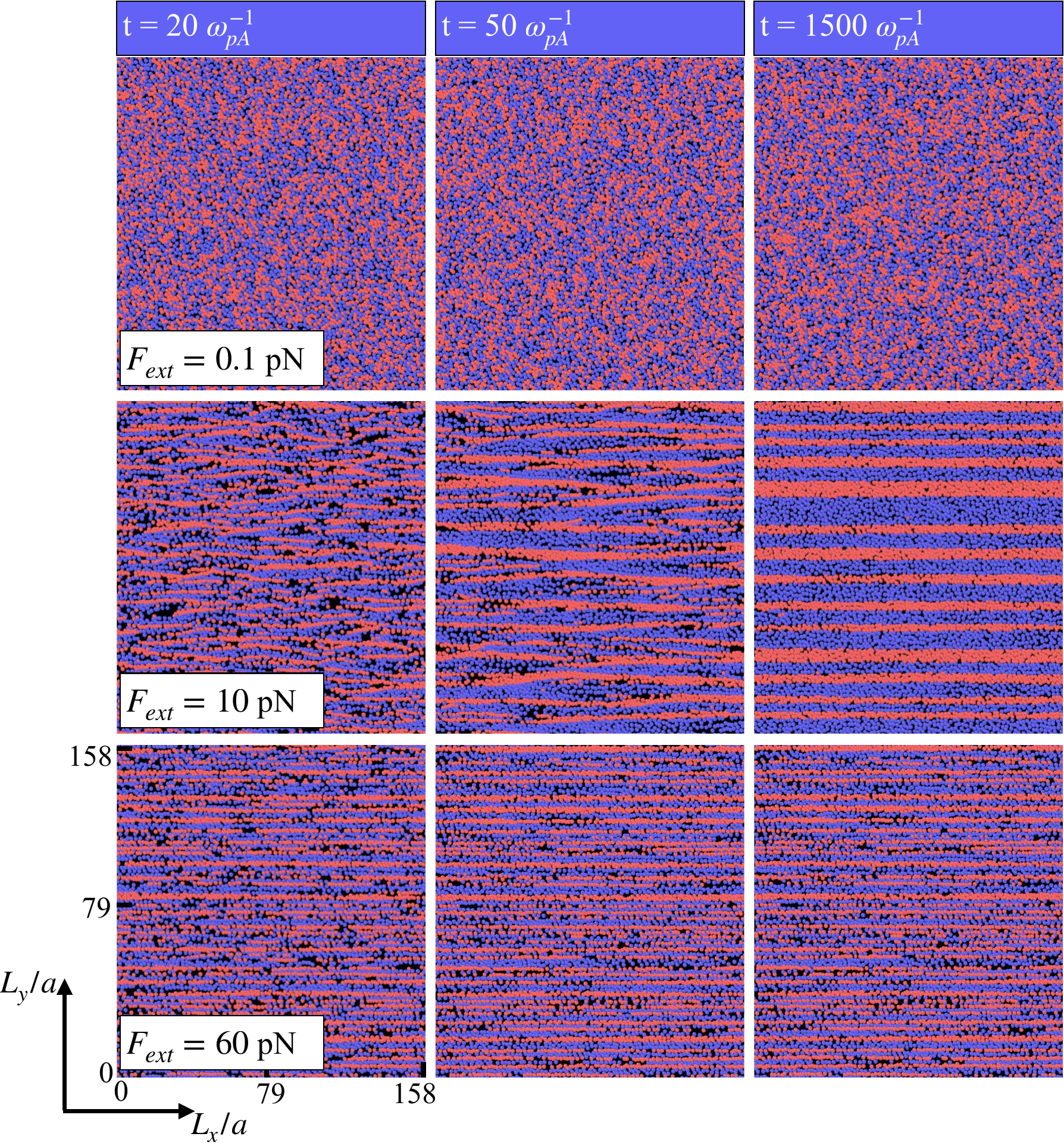}
\caption{The effect of $F_{ext}$ on the lane/band formation for  $\Gamma_A = 10$ and $\kappa = 0.9$ at different times.
With increasing force, we can see that there is a transition from random
distribution (first row) to band (second row) and then to formation of
lane (third row). The system size in both directions is $L_x/a = L_y/a = 158$.}
\label{fig:fig4}
\end{figure}
\paragraph*{}
 We show the time evolution of the critical binary mixture for different forcing amplitudes in Fig.~\ref{fig:fig4}. At low forcing ($F_{ext} = 0.1$ pN), there is no lane formation throughout the evolution, as can be seen from the first row. For moderate forcing ($F_{ext} = 10$ pN), initially, start to form lanes. At later times, due to finite cross diffusion, different branches try to merge and form bands in the steady state. There is a critical forcing ($F_{cri}$) for each value of coupling strength below which no lane and band forms. Within this forcing range of $F_{cri}$, thicker bands form as there is higher cross-diffusion perpendicular to the forcing direction. Lanes form for large forcing ($F_{ext} \approx 60$ pN) due to strong suppression of cross-diffusion. We find similar observations for $F_{ext}$ on the segregation of an off-critical mixture. The details, along with the figure, are provided in Appendix A. 
\paragraph*{}
 Figure~\ref{fig:fig5}, through its three rows, illustrates the time evolution of the critical binary mixture for different coupling strengths at a fixed $F_{ext}$ value. No lanes form (Row 1) for $\Gamma_A = 1$, even at a moderate forcing of $F_{ext} = 5$ pN. For $\Gamma_A = 10$, a mixture of lanes and bands emerges (Row 2). For $\Gamma_A = 100$, both lanes and bands are observed again, with a predominance of the band. Additionally, the structures appear ordered for higher value of the coupling strength. This is due to the fact that $v_{th}$ decreases with increasing coupling strength. Furthermore, a detailed analysis of lane and band formation in the binary dusty plasmas by calculating various physical quantities is presented in the following subsections.
 
\begin{figure}[t]
\includegraphics[width=0.5\textwidth]{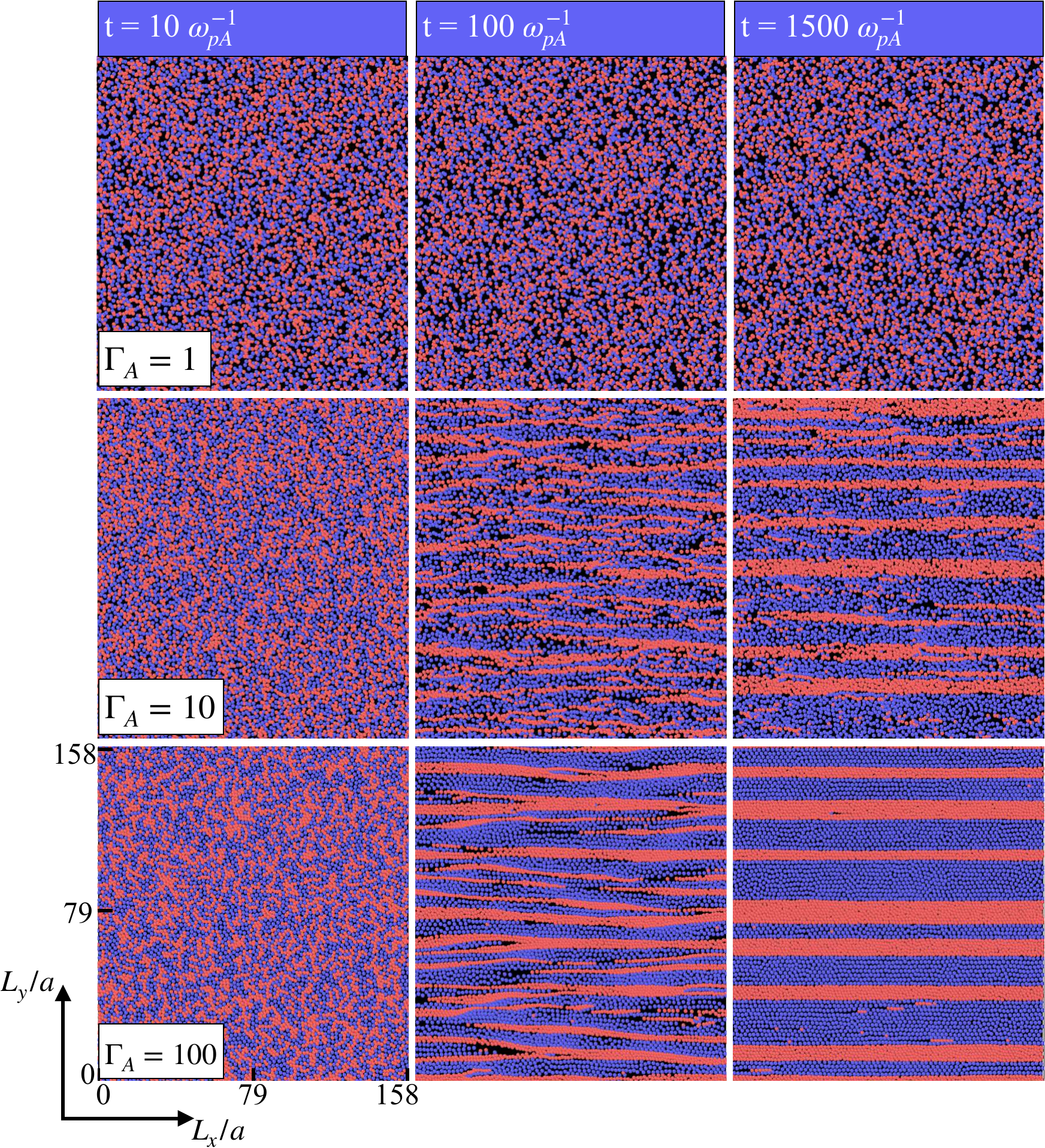}
\caption{The effect of the coupling strength on the lane/band formation at
different times. The value of $F_{ext} = 5$ pN and the $\kappa = 0.9$. The system size in both directions is $L_x/a = L_y/a = 158$.}
\label{fig:fig5}
\end{figure}
\subsection{The effect of $F_{ext}$ on the coupling strength, $\Gamma_f/\Gamma_A$}
\label{subsec:coupling_strength}
\begin{figure}[ht!]
\includegraphics[width=\linewidth]{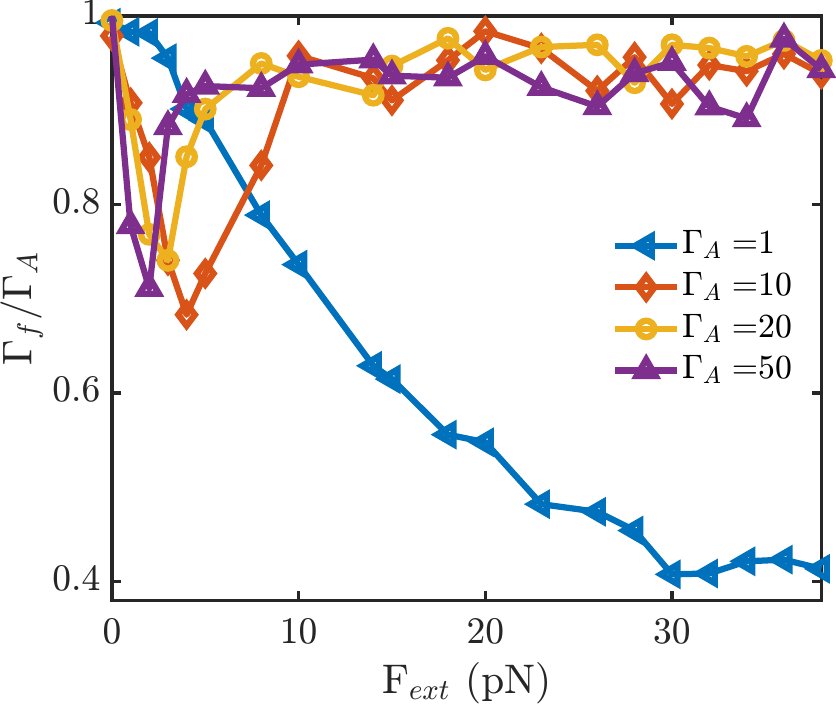}
\caption{The variation in the final to initial coupling strength ratio with increasing external forcing. $\kappa$  is 0.9 and  $ n_{dr} = 3$ for all cases.}
\label{fig:fig6}
\end{figure}

The external forcing amplitude is chosen such that the system temperature does not change much with $F_{ext}$. The reason behind this is that if the temperature remains constant, then we can see the effect of $\Gamma$ on the formation and dynamics of the lanes. Figure~\ref{fig:fig6} shows the relative change in the coupling strength (temperature) with different forcing amplitudes. Notably, the temperature of system rises only for a narrow range of initial forcing leading to a dip in the $\Gamma_f/\Gamma_A$ profile and does not change significantly for large values of external forcing amplitude. We calculate it at $t = 400~\omega_{pA}^{-1}$ where the temperature almost saturates.
Here, $\Gamma_f$ represents the coupling strength in the steady state. In the direction of the applied forcing, the center-of-mass velocity ($v_{com}$) is subtracted from each particle's velocity ($v_i$) to obtain the thermal velocity ($v_{th} = v_i-v_{com}$) from which temperature is calculated and hence $\Gamma_f$. $\Gamma_{A}$ is the coupling strength corresponding to its initial temperature.
\paragraph*{}
We explain the observed trend of $\Gamma_f/\Gamma_A$ as a function of the external force $F_{ext}$ as follows:
In the absence of an external force, particle motion is purely thermal, and the system’s velocity corresponds to the thermal velocity, $v_{th}$. With the application of an external force, collisions are enhanced, which is associated with an increase in the kinetic energy. 
The increase in overall kinetic energy leads to an effective velocity gain, which we denote as $v_c$. The resulting relative velocity can then be expressed as $v_{rel} = v_{th} + v_{c}$.   
\paragraph*{}
As the external force increases, the collision frequency increases, leading to a further increase in $v_c$ and, consequently, the kinetic temperature of the system. This
increase in temperature reduces the effective coupling strength, which is observed in the form of a drop in the value of $\Gamma_f/\Gamma_A$, see fall trend in the curves of Fig.~\ref{fig:fig6}.
With further increase in $F_{ext}$, particles begin to align along the direction of the force and start forming lanes. This structural ordering reduces the frequency of interparticle collisions and the system becomes more ordered. As a result, the temperature starts to decrease, and $\Gamma_f/\Gamma_A$ begins to rise again, which leads to V-shape curves in Fig.~\ref{fig:fig6}. Moreover, at sufficiently high forcing, collisions become rare and the contribution of $v_c$ diminishes. The relative
velocity approaches the thermal value again i.e. $v_{rel} \approx v_{th}$, and accordingly $\Gamma_f/\Gamma_A$ attains values close to its equilibrium value.
\paragraph*{}
 Since collision frequency increases with $\Gamma_A$~\citep{Baalrud_2015_PRE}, consequently energy redistribution becomes more efficient at lower values of $F_{ext}$, leading to faster thermalization. Thus, the formation of the structure occurs early in the system for the higher value of coupling strength, leading to a leftward shift in the minima of $\Gamma_f/\Gamma_A$ with increasing coupling strength (see Fig.~\ref{fig:fig6}).
\paragraph*{}
We have also studied the influence of the driving frequency $f_{dr}$ on the structure formation in the system. Our observations reveal a transition from band to lane and eventually to no structure formation as $f_{dr}$ decreases. This is provided in Table~\ref{tab:table2}, where the first column represents the applied driving frequencies, while the second column indicates the corresponding structure formation observed in the system. The transition reflects a shift from the smooth to the discrete forcing regime as particles get sufficient time to diffuse between successive regimes.
\begin{table}[h!]
{\renewcommand{\arraystretch}{1.9}
\caption{Conditions for lane/band formation under varying forcing frequency $\omega_{dr}$ at a fixed amplitude of $F_{\text{ext}} = 30$ pN.}
\begin{center}
\begin{tabular}{l @{\hspace{3cm}} l}
\hline
\hline
Case  & Structure \\
\hline
$\omega_{dr} >> \omega_{pA}$& Lane  \\
$\omega_{dr} > \omega_{pA}$& Band \\ 
$\omega_{dr} = \omega_{pA}$& No structures\\ 
$\omega_{dr} < \omega_{pA}$& No structures \\ 
\hline
\hline
\end{tabular}
\label{tab:table2}
\end{center}
}
\end{table}
\subsection{Order parameter for critical and off-critical mixtures}
The order parameter ($\phi_0$) is used to identify phase separation in binary systems~\citep{Dzubiella_2002_PRE, Sarma_2020_PoP}. We calculate it using the following expression:
 \begin{equation}
 \phi_0 = \left< \frac{1}{N_k} \sum_{i=1}^{N_k} \left|\frac{n_A^i - n_B^i}{n_A^i + n_B^i} \right| \right>,
    \label{order_para}
\end{equation}
where $n_{A,B}^i$ represents the number of particles of each species in a given bin and $N_k$ is the total number of bins. To calculate $\phi_0$, the simulation region is divided into $N_k$  bins along the direction perpendicular to the forcing. The bin width is set to approximate $5/7a$ in the present case to ensure sufficient particles of each species for improved statistics. Bins with no particles are rare and have been omitted from the calculation. 
\begin{figure}[b]
\includegraphics[width=0.9\linewidth]{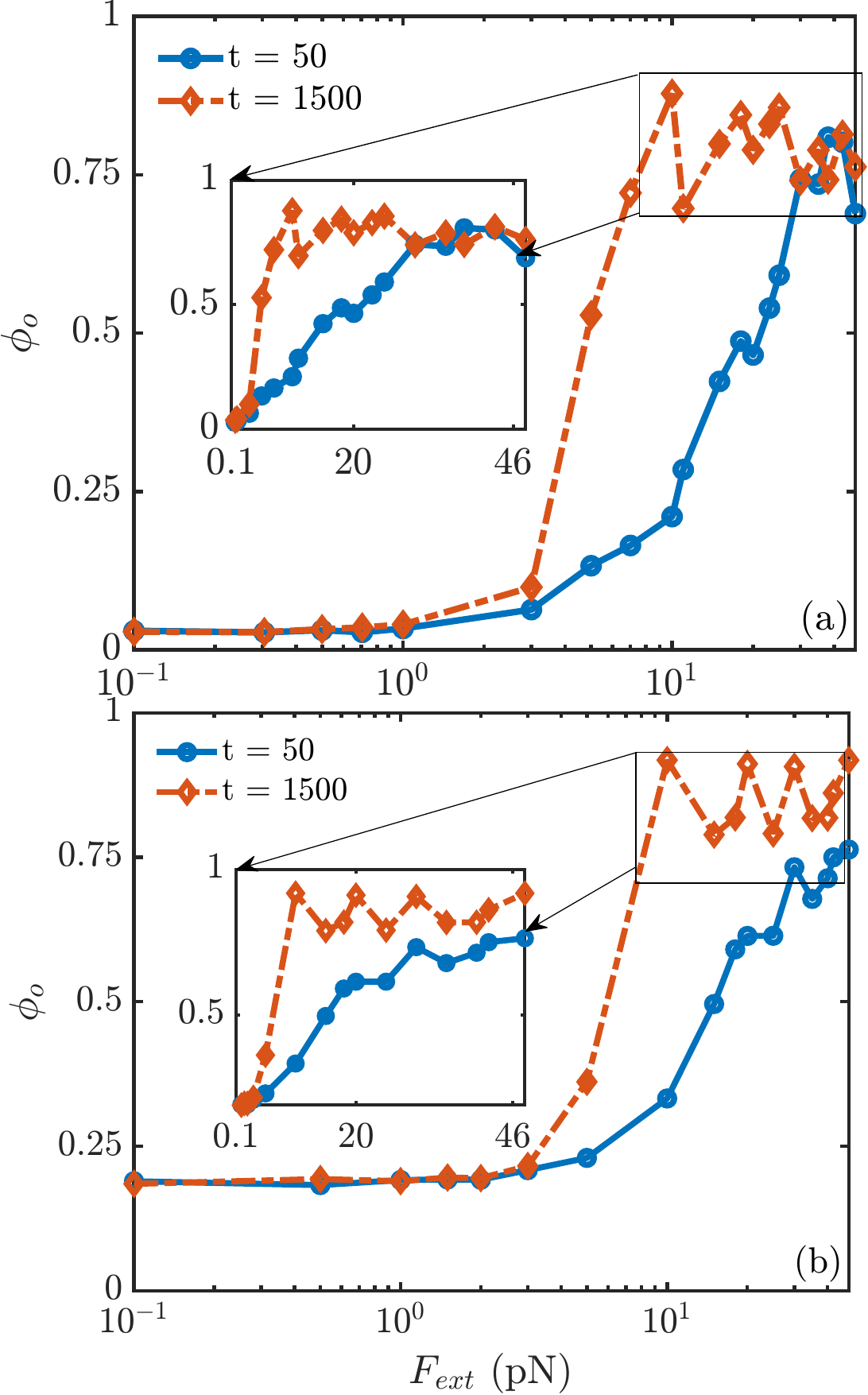}
\caption{Order parameter variation with the external forcing for (a) critical and (b) off-critical ($\Psi_A:\Psi_B \vert$ 0.3:0.7) mixtures. The value of $\Gamma_A = 10$ and the $\kappa =0.9$ for both the subplots. }
\label{fig:fig7}
\end{figure}
\begin{figure*}[t]
\includegraphics[width=\textwidth]{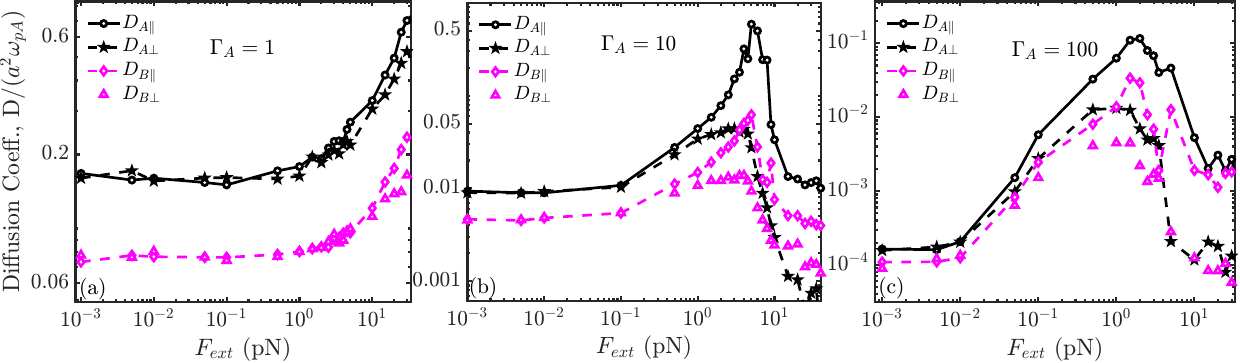}
\caption{The effect of coupling strength on $D_{\parallel}$ and $D_{\perp}$ for $\Gamma_A $= 1, 10, and 100. The screening parameter $\kappa$ is 0.9, and $\Psi_A = 0.5$.  }
\label{fig:fig8}
\end{figure*}
\paragraph*{}
The order parameter $\phi_0$ as a function of the $F_{ext}$ amplitude for (a) critical mixture and (b) off-critical mixture is shown in Fig.~\ref{fig:fig7}. The concentration ratio of type A and type B species for critical mixture and off-critical mixture are 0.5:0.5 and 0.3:0.7, respectively. The value of $\phi_0$ is relatively small for small forcing amplitudes as the system is randomly mixed. As $F_{ext}$ increases, it induces species segregation, leading to a gradual rise in the $\phi_0$. Eventually, it reaches a saturated state with a maximum value of $\phi_0$ approximately 1, indicating the formation of lanes and bands, as shown in Fig.~\ref{fig:fig7}. With an increase in the $F_{ext}$, the parallel component of velocity increases, resulting in the motion of particles in a perpendicular direction getting restricted. This dynamic behavior favors the formation of lanes over bands at higher forcing strengths. In the intermediate range of force, band formation is favorable due to the significant motion of particles in both parallel (drift) and perpendicular (diffusion) directions. The competition between particle flow in parallel and perpendicular directions determines the formation of mixtures, lanes, and bands in the medium.
\paragraph*{}
The evolution of the order parameter over time provides insight into the onset and dynamical progression of lane and band formation in binary complex plasma. Figure~\ref{fig:fig7} displays that $\phi_0$ with increase in $F_{\text{ext}}$ becomes steeper and more step-like in the steady state ($t = 1500~\omega_{pA}^{-1}$) compared to its smoother form at earlier times ($t = 50~\omega_{pA}^{-1}$). The insets in both plots show the saturation of $\phi_0$ at higher forcing values. 
\paragraph*{}
Although the difference in the order parameter between the critical and off-critical mixtures is small, we observe that at very low values of forcing, the off-critical mixture exhibits higher order compared to the critical mixture. This increased order in the off-critical case arises from the higher concentration of type B particles, which have a greater coupling strength and, consequently, a higher degree of structural order.
\subsection{Diffusion coefficient in the critical mixture}
In any gaseous or liquid medium, the diffusion coefficient (D) quantifies the rate at which particles diffuse over time due to random thermal motion. It is calculated using the mean-square displacement of particles:
 \begin{equation}
 D = \lim_{t \to \infty} \frac{1}{4t} \left\langle \frac{1}{N} \sum_{i=1}^{N} \left[ \left( \mathbf{r_i(t) - r_i(0)} \right) \cdot \hat{r} \right]^2 \right\rangle. 
    \label{Diff_coeff2D}
\end{equation}
Here $\mathbf{r_i(t)}$ is the position of $i$-th particle at time $t$. In the calculations, the mean squared displacement (MSD) is averaged over the trajectories of the particles. 
\paragraph*{}
When an external force is applied or the particles experience drift in a particular direction, the diffusion coefficient becomes directional and dependent on the force. For instance, in a magnetized strongly coupled plasma, the diffusion coefficients along and perpendicular to the magnetic field are distinct, see ~\citet{Ott_2011_PRL}. Similarly, the diffusion coefficients exhibit anisotropy in lane formation processes due to the external drive~\citep{Dzubiella_2002_PRE}. To isolate the purely diffusive behavior from the net drift, we calculate the diffusion coefficients parallel ($D_\parallel$) and perpendicular ($D_\perp$) to the direction of the external force following the approach in~\citet{Vissers_2011_SM}:
\begin{equation}
 D_{\perp, \parallel} = \lim_{t \to \infty} \frac{1}{2t}
 \left\langle \frac{1}{N} \sum_{i=1}^{N}
 \left[ \left( \Delta \mathbf{r_i}(t) - \left\langle \Delta \mathbf{r}(t) \right\rangle \right) \cdot \hat{r}_{\perp,\parallel} \right]^2
 \right\rangle, 
 \label{Diff_coeff1D}
 \end{equation}
where
 \[ \Delta \mathbf{r}_i(t) = \mathbf{r}_i(t) - \mathbf{r}_i(0) \hspace{0.1cm} \text{and} \hspace{0.1cm} \left\langle \Delta \mathbf{r}(t) \right\rangle = \frac{1}{N} \sum_{i=1}^{N} \Delta \mathbf{r}_i(t) .\] This ensures that the contribution of directed motion is subtracted out. Since the force is applied to type A particles, inter-particle scattering is expected to influence $D_{\parallel,\perp}$ of type B as well. Furthermore, the strong coupling also influences the diffusion coefficient by altering the thermal motion of the particles.
 \paragraph*{}
 The components of the diffusion coefficient for both particle types as a function of the external force $F_{ext}$ are plotted in Fig.~\ref{fig:fig8}, which includes three subplots: (a), (b), and (c), corresponding to coupling strengths $\Gamma_A = 1, 10$ and $100$, respectively. 
The behavior of the diffusion coefficient can be divided into three distinct regimes: an initial constant stage, a rising phase, and a final decreasing phase after reaching a maximum. In the initial regime, the diffusion coefficient remains nearly constant due to the application of a very small external force. This weak forcing induces negligible drift velocity, resulting in minimal drift-driven scattering among particles and, consequently, an insignificant effect on the diffusion.
 As force further increases, it creates significant drift among type A particles, which leads to enhancement in the inter-species scattering and diffusion (see the rising region of Figure~\ref{fig:fig8}).
At higher forcing, particles start forming lanes/bands, resulting in interspecies scattering getting reduced, which leads to reduced diffusion coefficient along the perpendicular and parallel directions (see the fall region in the subplots (b) and (c) of Figure~\ref{fig:fig8}).
\paragraph*{}
In the strongly coupled case ($\Gamma_A = 100$), the particles have reduced thermal motion, allowing them to align and form lanes or bands under relatively low external forcing. As a result, lane formation occurs at lower force magnitudes, which is reflected in the characteristic peak and subsequent drop in the diffusion coefficient represented by subplot (c) of Figure~\ref{fig:fig8}).
In contrast, for lower coupling strengths, such as $\Gamma_A = 1$, thermal motion is more significant, and a stronger external force is required to induce lane formation. This is reflected in subplot (a) of Figure~\ref{fig:fig8}), where no clear peak or drop in the diffusion coefficient is observed, indicating the absence of lane or band formation within the examined force range.  
\paragraph*{}
Finally, a clear difference in the diffusion coefficients of type A and type B particles is observed across all subplots. This disparity arises from the variation in their coupling strengths due to differences in their charges. Particles with higher coupling strengths experience stronger interparticle correlations and reduced thermal motion, leading to lower diffusion coefficients \citep{Hou_2009_PRL}. 
\subsection{Finding critical force value, $F_{cri}$ } 
\paragraph*{}
So far, we have studied how the binary mixture evolves at different values of  $F_{\text{ext}}$ and how this evolution is reflected in the drift velocity ($v_d$), the coupling strength ($\Gamma_f/\Gamma_A$), the diffusion coefficient ($D_{\parallel,\perp}$), and the order parameter ($\phi_0$).
\begin{table}[ht!]
{\renewcommand{\arraystretch}{1.5}
\begin{center}
\caption{The value of the critical force depends on the method of calculation.}
\begin{tabular}{l@{\hspace{1.2cm}} l@{\hspace{1.2cm}} l}
\hline
\hline
 Figure & Approaches  & $F_{cri}$ \\
\hline
Fig.~\ref{fig:fig3} & Drift velocity ($v_d$) &  1 pN \\
Fig.~\ref{fig:fig6} & Coupling strength ($\Gamma_f/\Gamma_A$) & 4 pN\\
Fig.~\ref{fig:fig8} &Diffusion coefficient ($D_{A\perp}$)  & 3 pN\\
Fig.~\ref{fig:fig7}& Order parameter ($\phi_0$) & 5 pN  \\
\hline
\hline
\end{tabular}
\label{tab:table3}
\end{center}}
\end{table}
Notably, we can tentatively locate the $F_{cri}$ value independently from all these diagnostics, and there is a profound similarity in the value.
The approximate values of $F_{cri}$ based on different calculation methods are summarized in Table~\ref{tab:table3}.
We analyzed the effect of $F_{ext}$ on drift velocity; the crossover of the $v_d$ and $v_{th}$ provides the $F_{cri}$. Our observations suggest that lanes/bands form only when $v_d$ dominates $v_{th}$.
In Fig.~\ref{fig:fig6}, the ratio $\Gamma_f/\Gamma_A$ exhibits a dip as a function of $F_{ext}$. This dip defines $F_{cri}$, and the corresponding value is approximately 4 pN. Now, if we take the case of ($D_{A\perp}$) with external forcing, it attains a maxima above which it starts decreasing. The $F_{ext}$ value at which $D_{A\perp}$ reaches its maximum defines the critical force, which is approximately 3 pN.
The $F_{cri}$ value calculated from the order parameter is slightly higher compared to the other three methods, which is around 5 pN, and it corresponds to $\phi_0 = 0.5$.
\begin{figure}[b]
\includegraphics[width=0.8\linewidth]{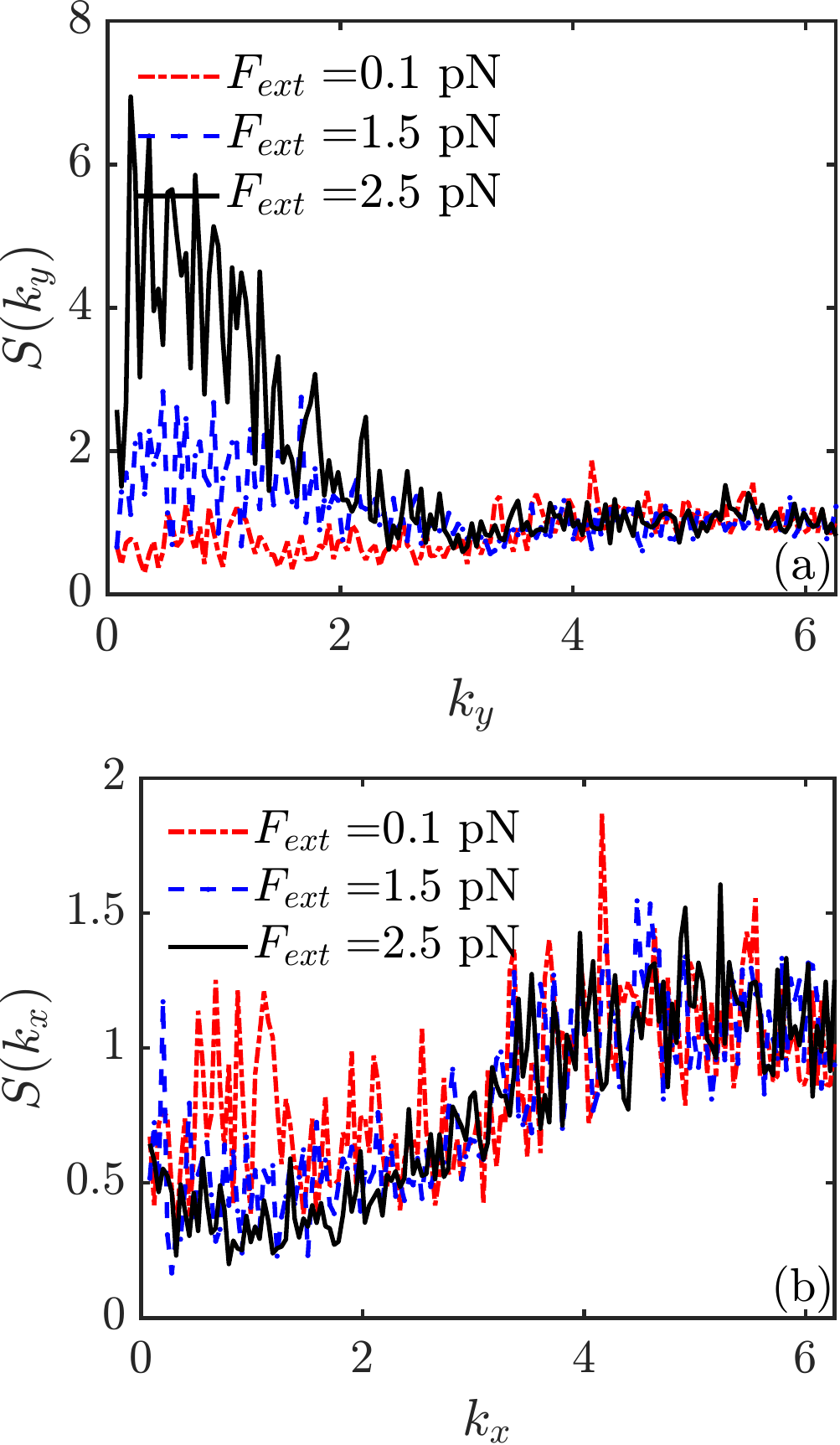}
\caption{Observation of (a) low-$k_y$ peaks in the $S(k_y)$ and (b) formation of low-$k_x$ dips in the $S(k_x)$. This representation is the structural inhomogeneity in the system. Both the plots are calculated at $t = 500~\omega_{pA}^{-1}$.}
\label{fig:fig9}
\end{figure}
\subsection{Scaling in critical and off-critical mixtures}
\label{subsec:scaling}
\paragraph*{}
For a given species, we calculate the average size of lane/band using structure factor $S(k,t)$ which reads~\citep{Farida_2024_PoP,Gidituri_JCP_2019,Wysocki_2009_PRE}
 \begin{equation}
S(k,t) = \frac{1}{N} \left[\left(\sum_{i=1}^{N}\cos(\mathbf{k} \cdot \mathbf{r}_{i})\right)^2 + \left(\sum_{i=1}^{N}\sin(\mathbf{k} \cdot \mathbf{r}_{i})\right)^2 \right],
\label{str_factor}
\end{equation}
where $ \mathbf{k} = 2\pi/{L} (n_x \hat{i} + n_y \hat{j})$, $n_x, n_y = \pm 0,\pm 1,....., \pm n_{\text{cut}}$. 
We have calculated the parallel ($S(k_x,t)$) and perpendicular ($S(k_y,t)$)  components of the structure factor for both the mixtures using Eq.~\ref{str_factor}. Figure~\ref{fig:fig9} with subplots (a) and (b) representing the parallel and the perpendicular component, respectively, for the critical mixture. 
Both the subplots show a flat peak in structure factors $S(k_x)$ and $S(k_y)$ saturated at a value one. This saturation is a usual depiction of homogeneity in the medium. This peak is referred as the ``characteristic". As lanes and bands have width in the Y-direction, subplot (a) reflects a few ``additional peaks" at low values of $k_y$ corresponding to the width of lane/band. We can also observe that the ``additional" peak height increases with increasing forcing, showing an increase in the formation of the lane/band with forcing. We calculate both components of the structure factor when the system reaches steady state.
\begin{figure}[b]
\includegraphics[width=0.8\linewidth]{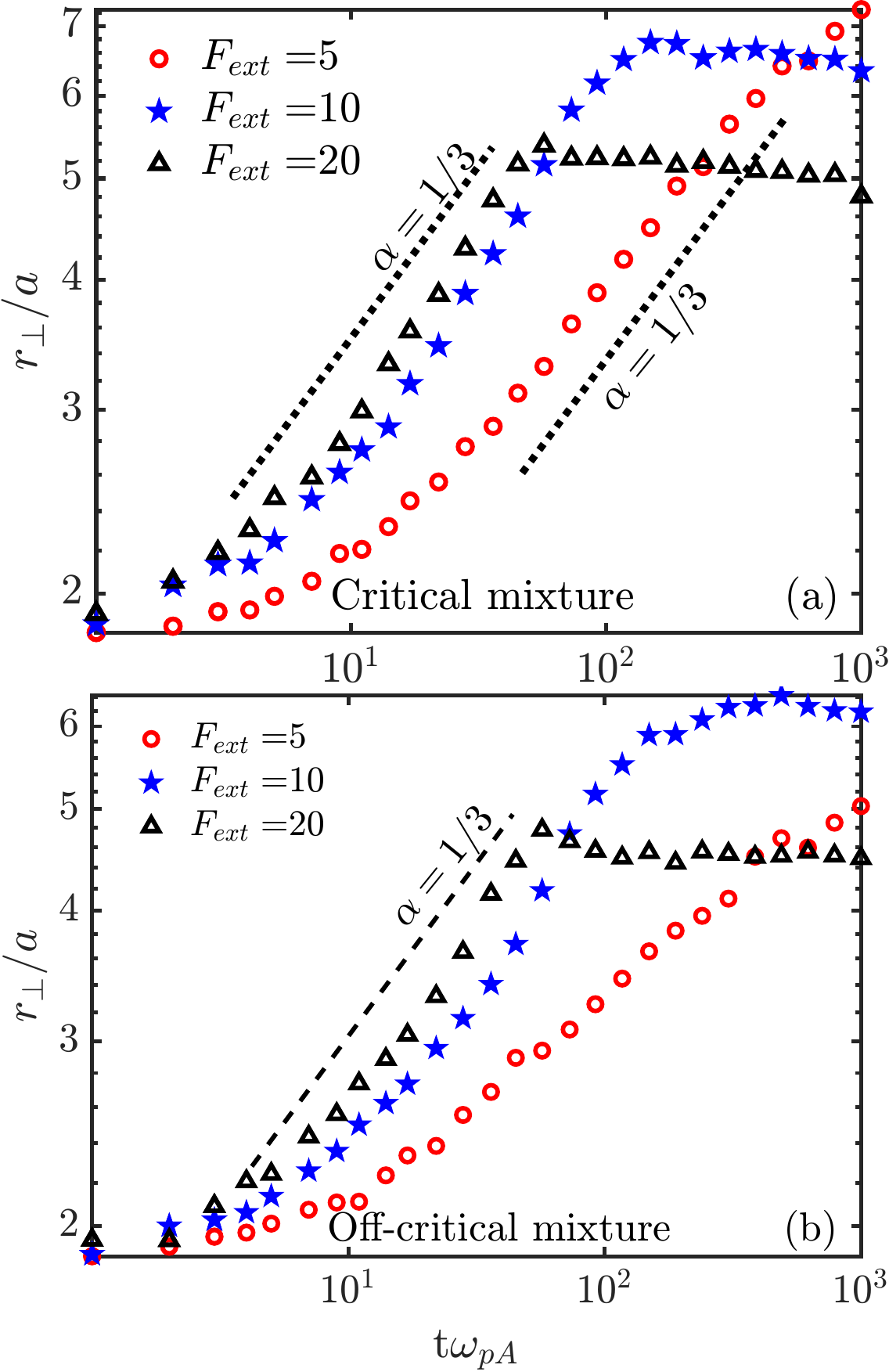}
\caption{The lane and the band's average size ($r_{\perp}$) as a function of time in the perpendicular direction of $F_{ext}$ for (a) critical and (b) off-critical mixtures. The $\alpha$ slope lines are for reference only and not the fit to data parameters.}
\label{fig:fig10}
\end{figure}
\paragraph*{}
For forces greater than $F_{cri}$ bands or lanes form. Though the formation of such structures evolves with time and attains a steady state, during this evolution, the typical width of lanes or bands follows a scaling. We estimate the typical lane/band size evolution represented by Fig.~\ref{fig:fig10} for the critical subplot (a) and off-critical subplot (b) mixture. As the structures align with the force direction, we measure the change in average width $r_{\perp}$ in the direction perpendicular to the forcing.
\paragraph*{}
The characteristic length of the lane/band in the direction perpendicular to the forcing is given by
\begin{equation}
     r_{\perp}(t) = 2\pi\frac{\sum_{k=0}^{kcut} S(k_y,t)}{\sum_{k=0}^{kcut} k_yS(k_y,t)},
\label{domain_size}
\end{equation}
in which $S(k_y,t)$ is the perpendicular component of the structure factor and $k_y$ is the wave vector along the Y-direction. In Eq.~\ref{domain_size}, $k_{cut} = 2\pi/a$, where $a$ is the average inter-particle separation, the smallest relevant length scale in the modeled system. The lane and band sizes are calculated using Eq.~\ref{domain_size} with this choice of $k_{cut}$ and later compared and validated with the average lane and band sizes in the actual simulation snapshots. 
The width of the band increases with time and eventually saturates at a maximum value. We consider three cases to compare this behavior under different driving forces: $F_{ext} = 5, 10$, and 20 pN.
For small force ($F_{ext} = 5$ pN), as shown in subplot (a) of Fig.~\ref{fig:fig10} (red circles), the size $r_{\perp}(t)$ grows with time but does not reach saturation within the observed timescale. However, at higher forces ($F_{ext} =$ 10 and 20 pN), $r_{\perp}(t)$ increases initially, reaches a peak, and then saturates. Notably, the saturation value at $F_{ext} =20~$pN (represented by black triangles) is lower than that at $F_{ext} = 10$ pN (blue stars). This occurs because increasing the external force leads to a transition from bands to lanes. During the growth phase, the size $r_{\perp}(t)$ follows a power-law scaling with time, given by $r_{\perp}(t) \propto t^{\alpha}$, where $\alpha$ is the scaling exponent. We find that the value of $\alpha$ is approximately 1/3. 
The $r_{\perp}(t)$ in the off-critical mixture exhibits the identical behavior as the critical mixture for different force values, as shown in subplot (b) of Fig.~\ref{fig:fig10}. In this case too, the size $r_{\perp}(t)$ follows a power-law growth with time, with the exponent $\alpha~\approx$ 1/3.
\paragraph*{}
We note that the size of the phase-separated domains has not been systematically reported or analyzed in existing simulations or experiments involving dusty plasmas under external driving. This lack of data limits the possibility of a direct comparison with previous studies related to dusty plasma.
However, previous studies on equilibrium phase separation have reported a domain growth exponent of 1/2 for critical mixtures~\citep{Farida_2024_PoP}. In contrast, our results show a growth exponent of approximately 1/3 for the evolution of lane or band structures.
This deviation can be attributed to two factors: the viscoelastic behavior of dusty plasmas at moderate to high coupling strengths ($\Gamma$)~\citep{Feng_PRL_2010} and the continuous external driving applied in our system. The latter keeps the system out of equilibrium and significantly modifies the coarsening dynamics, leading to a distinct scaling regime. 
\subsection{Effect of coupling strength and the concentration on the drift velocity}
\label{subsec:drift_gamma_phi}
\begin{figure}[t]
\includegraphics[width=0.9\linewidth]{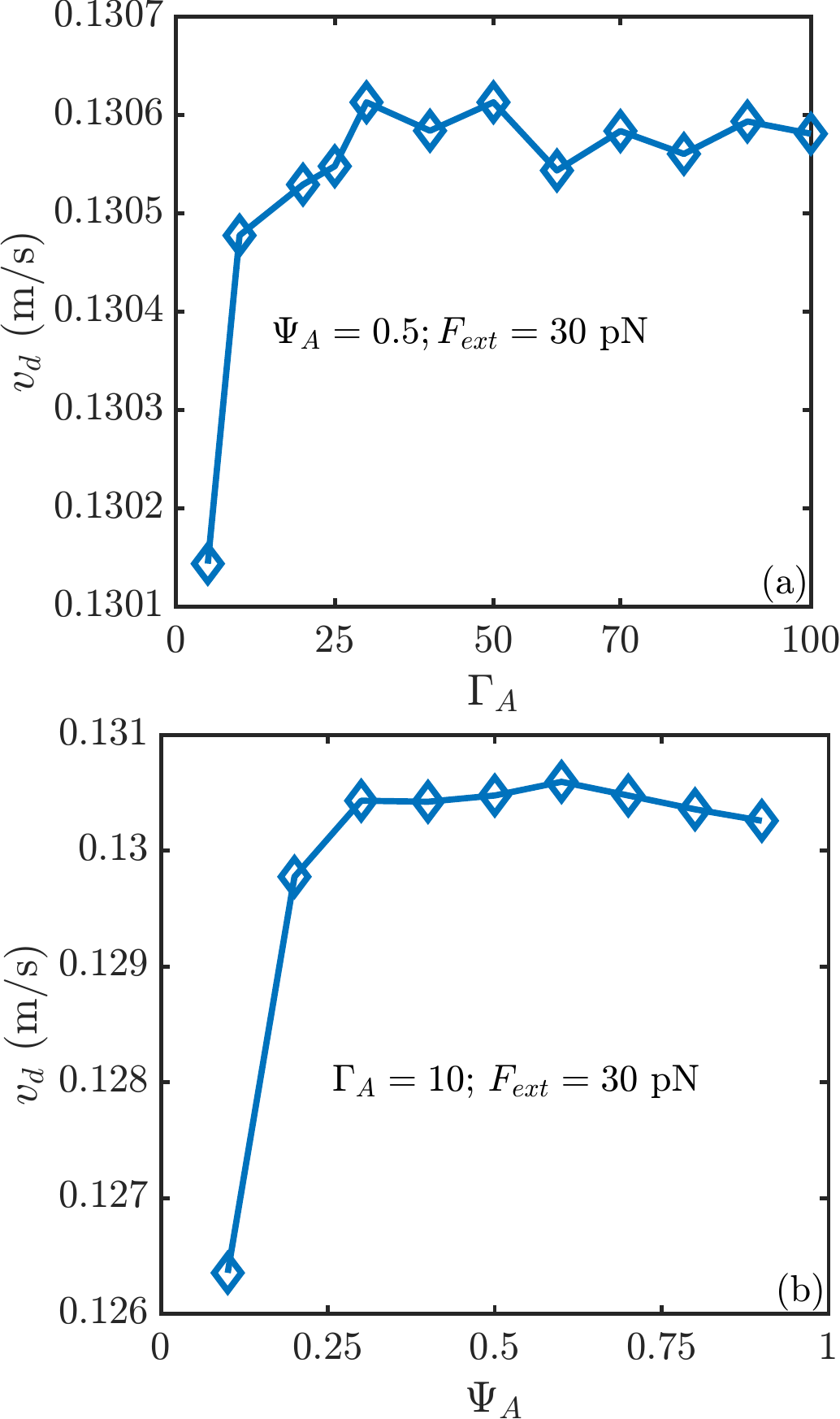}\caption{(a) The effect of the coupling strength on the drift velocity is calculated in the steady state. The value of $F_{ext} = 30$ pN and $\Psi_A = 0.5$. (b) The effect of the concentration on the drift velocity for $\Gamma_A = 10$ and $F_{ext} = 30$ pN.}
\label{fig:fig11}
\end{figure}

\paragraph*{}
Our observations suggest that species A attains a particular drift velocity for each case of band and lane formation. Figure~\ref{fig:fig11} shows the effect of $\Gamma_A$ and $\Psi_A$ on the amplitude of drift velocity (for a particular $F_{ext}$) in subplots (a) and (b), respectively. From subplot (a), the variation of $v_d$ shows that an increase in $\Gamma_A$ at small values leads to sharp increase in $v_d$. The drift velocity $v_d$ exhibits a slight rise from approximately 0.1301~m/s to 0.1306~m/s. Although the absolute change is small (less than 0.4\%), this trend suggests a gradual enhancement in directed motion as the system becomes more strongly coupled. This behavior can be attributed to decrease in particle kinetic energy with increasing $\Gamma_A$, which leads to reduced thermal velocity. As a result, thermal fluctuations become less effective in counteracting the applied external force, allowing a larger fraction of the force to contribute to net drift $v_d$.
However, when comparing subplot (a) with Fig.~\ref{fig:fig3}, it is evident that the external force has a far more pronounced effect on $v_d$ than changes in coupling strength.
\paragraph*{}
Furthermore, for the same $F_{ext}$ as in subplot (a), we now compute $v_d$ while increasing composition $\Psi_A$, as shown in subplot (b). We observe that when $\Psi_A$ is small (up to $\approx 0.3$), $v_d$ increases. However, beyond this value, $v_d$ nearly saturates. This behavior reflects the influence of type B on type A. When $\Psi_A$ is small, the heavier particles (type B) that are more strongly coupled dominate, significantly affecting type A. As $\Psi_A$ increases, this effect weakens and becomes negligible at higher concentrations of type A, leading to saturation in $v_d$.
\section{Discussion}
\label{sec:discus}
\begin{figure}[ht!]
\includegraphics[width=\linewidth]{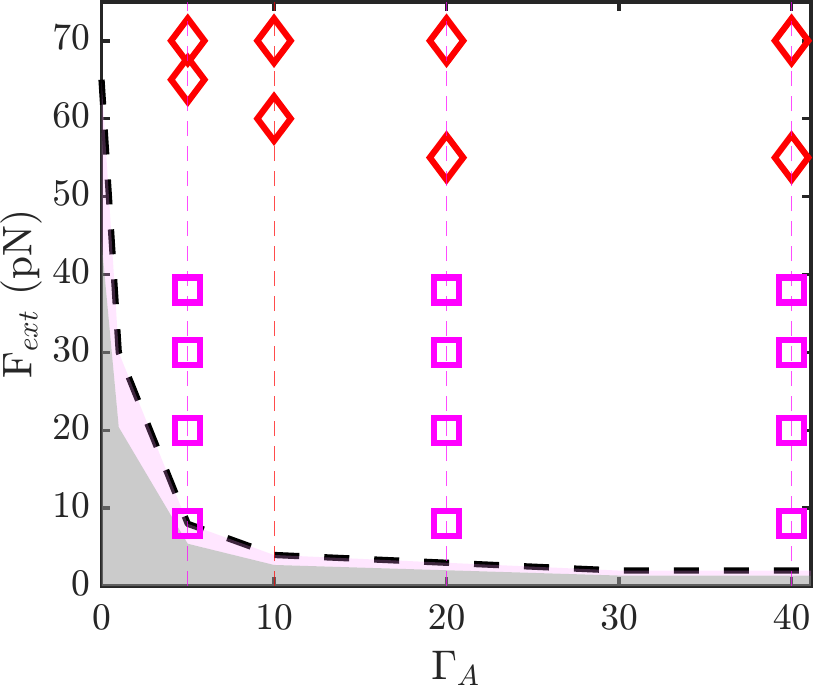}
\caption{The $\Gamma_A-F_{ext}$ phase diagram with regions of lane (diamond), band (square), and no structure formation (bottom shaded).}
\label{fig:fig12}
\end{figure}

\paragraph*{}
Our findings show that the critical force exhibits an exponentially decaying trend as a function of the coupling strength. We have also provided a phase diagram in the $\Gamma_{A}$ and the $F_{ext}$ space in Fig.~\ref{fig:fig12}. The black-dashed line is the critical force for different values of the coupling strength. The background patch (light magenta) represents simulation fluctuations range in the estimation of the critical force values. The grey-shaded region below this dashed line represents no structure formation. The magenta squares represent the band region, and the red diamonds represent the region of the lane formation. We can see from the figure that at very low $\Gamma_{A}<2$, the critical value of $F_{ext}$ is very large. Also, we can observe that the lane/band formation occurs at less force for higher $\Gamma_{A}$ values. The reason is that the thermal energy is higher at lower coupling strengths, requiring greater force to align the particles in the direction of the applied force. All simulation parameters remain the same for calculating this phase diagram, such as the $n$, $\gamma$, and the form of the forcing. An important observation is that while bands form just above the critical forcing, the formation of lanes in the steady state requires strong inhibition of cross-diffusion, which is only visible at very high values of $F_{ext}$. 
\paragraph*{}
Since the present work is based on simulation, we
have also look into experimental scenario which can be
used to create present study conditions in the binary
complex plasmas. 
In a previous experiment conducted by Liu et al.~\citep{Liu_2003_PoP}, melamine-formaldehyde (MF) particles of different sizes were accelerated using an Ar$^{+}$ laser in an RF discharge plasma. Their work calculated the coefficient $q$ (laser pressure radiation coefficient), which determines the efficiency of a particle's interaction with a laser beam. 
Using the same values used in the paper of~\citet{Liu_2003_PoP}, like $q$ and $\mu$ (refractive index), we provide a theoretical estimation of the laser intensity required to induce segregation for the particles used in our simulations. For this, we have used the value of the external force at which lane formation nearly initiates ($3$ pN) in our simulations to calculate the laser intensity ($I_{L}$) using the relation: $F_{L} = q \mu\pi I_{L} r_p^2/c$.  Here, $c$ is the velocity of the light, $F_{L}$ is the radiation pressure force, which we here equate to the $F_{ext}$, and $r_p$ is the radius of the particles. 
Since the external force is applied only to type A particles, we use the radius of type A particle to calculate $I_{L}$. Based on the simulation parameters, the required intensity is found to be $I_{L}=7.9457 \times 10^6$ Wm$^{-2}$. This value is approximately two orders of magnitude greater than the laser intensity reported in the experimental paper by~\citet{Liu_2003_PoP}.
\begin{table}[h!]
{\renewcommand{\arraystretch}{1.5}
\caption{Comparison between Colloids and Dusty Plasma}
\begin{tabular}{|p{1.8cm}|p{4.53cm}|p{2.0cm}|}
\hline
\textbf{Particulars}  &   \textbf{Colloids}&  \textbf{Dusty plasma }   \\
\hline
Size ($\sigma$) & 1nm - 1$\mu$m & 1-10 $\mu$m \\ 
\hline
 Charge ($z$) & $10^2 < z < 10^4 $ & $10^3 < z < 10^4$ \\ 
\hline
 \centering \vspace{0.2cm}$\psi$ & 
\begin{tabular}[t]{@{}l@{}}
$\psi<0.1$ (low) \\ 
$0.1<\psi<0.5$ (intermediate) \\ 
$\psi<0.74$ (solid or glassy state)
\end{tabular}
& $\psi<10^{-4}$ \\
\hline
\centering \vspace{0.2cm} $\kappa$ & 
\begin{tabular}[t]{@{}l@{}}
$\kappa = \sigma / \lambda = \infty$ (Hard sphere) \\ 
$\left. \begin{array}{l}
\kappa \equiv \sigma / \lambda \\ 
\kappa \ll 1
\end{array} \right\}$(Screened Coulomb)
\end{tabular} 
& $\kappa = a/\lambda < 10 $ \\
\hline
\end{tabular}
\label{tab:table4}
}
\end{table}
\paragraph*{}
The formation of lanes and bands is also a topic of significant interest in colloidal suspensions. The diagnostics used in this study are equally applicable to such systems, except that strong coupling effects are usually negligible. A charged colloidal suspension comprises colloidal particles suspended in the background of a highly concentrated molecular solvent containing also co-ions and counter-ions. 
The screening length ($\lambda$) and the diameter of the particle ($\sigma$) are two deciding length scales in colloids. These parameters, through the screening parameter ($\kappa = \sigma/\lambda$), decide the interaction strength between charged particles. Changing the screening parameter influences the interactions between colloidal particles, which may range from screened Coulomb interactions to hard-sphere-like behavior. Tuning $\kappa$ also changes the effective coupling strength ($\Gamma_{eff} = \Gamma \exp{(-\kappa)}$) of the charged colloidal suspension. Table~\ref{tab:table4} compares charged colloids with dusty plasmas and clearly suggests that strong coupling, an important parameter for dusty plasmas ($\Gamma \approx 1 - 1000$), is usually not relevant to colloids due to weak electrostatic interactions. The typical charge and screening length combinations in colloids result in significantly lower $\Gamma_{eff}$ values compared to those in dusty plasmas.
\paragraph*{}
Another critical parameter in colloidal suspensions is the packing fraction, defined as $\psi = N_c \pi \sigma^3/(6 V)$, where $N_c$ is the total number of particles, $V$ is the system volume~\cite{Pusey_1986_N, Hynninen_2003_PRE, Royall_2006_JCP}. The packing fraction $\psi$ determines the various phases of the colloidal suspension and can be categorized into three distinct regimes: dilute suspension with $\psi<0.1$, intermediate $0.1<\psi<0.5$, and solid or glassy state $<0.74$. The packing fraction is also essential for understanding the interaction strengths and, hence, the physical states of charged colloidal suspensions~\cite{Ivlev_2012_WSPC}. Unlike the colloidal suspension in complex plasma, the medium surrounding the dust particles is dilute. The size of the dust particles typically ranges from 1–10 $\mu$m. For a number density of $10^{9}$ m$^{-3}$, the volume fraction for this particle size range is extremely small ($\psi < 10^{-4}$). As a result, the dust particles interact through long-range screened Coulomb interactions. The coupling strength can control the interaction strength between charged dust particles, leading to the system's various phases. Thus, in addition to the screening parameter, the coupling strength is a critical parameter in dusty plasmas, governing various phases of the system~\citep{Ivlev_2012_WSPC}.
\paragraph*{}
The comparison highlights that although colloidal suspensions and dusty plasmas are governed by similar fundamental interactions, their physical regimes differ significantly. In particular, dusty plasmas provide access to strong coupling and long-range interactions. These conditions are typically difficult to access in colloidal systems due to short-range screening and strong viscous damping from the solvent. These unique features make dusty plasmas an exceptional platform for exploring non-equilibrium structure formation with full particle-level resolution. Consequently, our findings not only deepen the understanding of structure formation in dusty plasmas under the influence of external drive but also offer insights that help to bridge the gap between complex plasmas and soft matter systems.
\paragraph*{}
The lane formation observed in our system exhibits qualitative similarity to that reported in driven colloidal suspensions, where such behavior has been theoretically explained through dynamical instability~\cite{Chakrabarti_2003_IOP, Chakrabarti_2004_PRE}. In their studies, a bifurcation in the steady-state density field occurs when the external force exceeds a critical threshold. It leads to anisotropic lane structures aligned with the driving direction.
 Although a full and quantitative linear stability analysis is required, we will try to establish it in the coming future as we gain some confidence over the calculations. Thus, we have qualitatively examined the potential for a similar instability in our system. Simply rearranging the interparticle interaction used by Chakrabarti et al., Yukawa potential, it becomes mathematically equivalent to the shielded Coulomb interaction used in our model.
Furthermore, our computed structure factors, along and perpendicular to the drive, exhibit anisotropic features consistent with those used by Chakrabarti et al. to diagnose dynamical instability. Based on these similarities in interaction potential, driving mechanism, and structural response, the lane formation observed in our system may likely be a result of dynamical instability.
\section{Summary}
\label{sec:summary}
\paragraph*{}
This paper details the dynamical evolution of a binary mixture of dusty plasma under external forcing. The study employs classical molecular dynamics simulations, explicitly incorporating strong particle correlations—a key feature of dusty plasmas.
The results demonstrate the lane and band formation dynamics in binary dusty plasmas for both critical and off-critical mixtures. Multiple independent diagnostics are utilized, including the order parameter ($\phi_0$), the drift velocity ($v_d$), diffusion coefficients ($D_{\parallel,\perp}$), domain size ($r_\perp(t)$), and the final-to-initial coupling strength ratio ($\Gamma_f/\Gamma_A$) to study the dynamics in the system. These diagnostics collectively highlight the non-equilibrium phase segregation.
All diagnostics consistently predict a critical forcing threshold required for segregation. Additionally, we observe a transition from band to lane formation. This transition occurs at higher forcing amplitudes due to a strong suppression of particle diffusion in the direction perpendicular to the applied force. In this work, we also highlight possible mechanisms through which such arrangements can be achieved in a controlled manner in the experiments. Furthermore, our study provides new insights into lane and band formation in colloidal mixtures. 
\paragraph*{}
We also obtained the phase diagram of binary mixtures in the coupling strength and the external force space. Based on these two parameters, we can reasonably distinguish the region of segregation and non-segregation. This phase space also provides information on the critical force as a function of coupling strength required for segregation to occur.
\begin{acknowledgements}
The authors acknowledge the use of AGASTYA HPC for the present work.
FB thanks the University Grants Commission (UGC) India for the fellowship. SKT and SK acknowledge the support from ANRF grants CRG/2020/003653 and NPDF fellowship, respectively. 
\end{acknowledgements}
\section*{Appendix A: Lane formation in off-critical mixture}
\begin{figure}[b]
\includegraphics[width=0.8\linewidth]{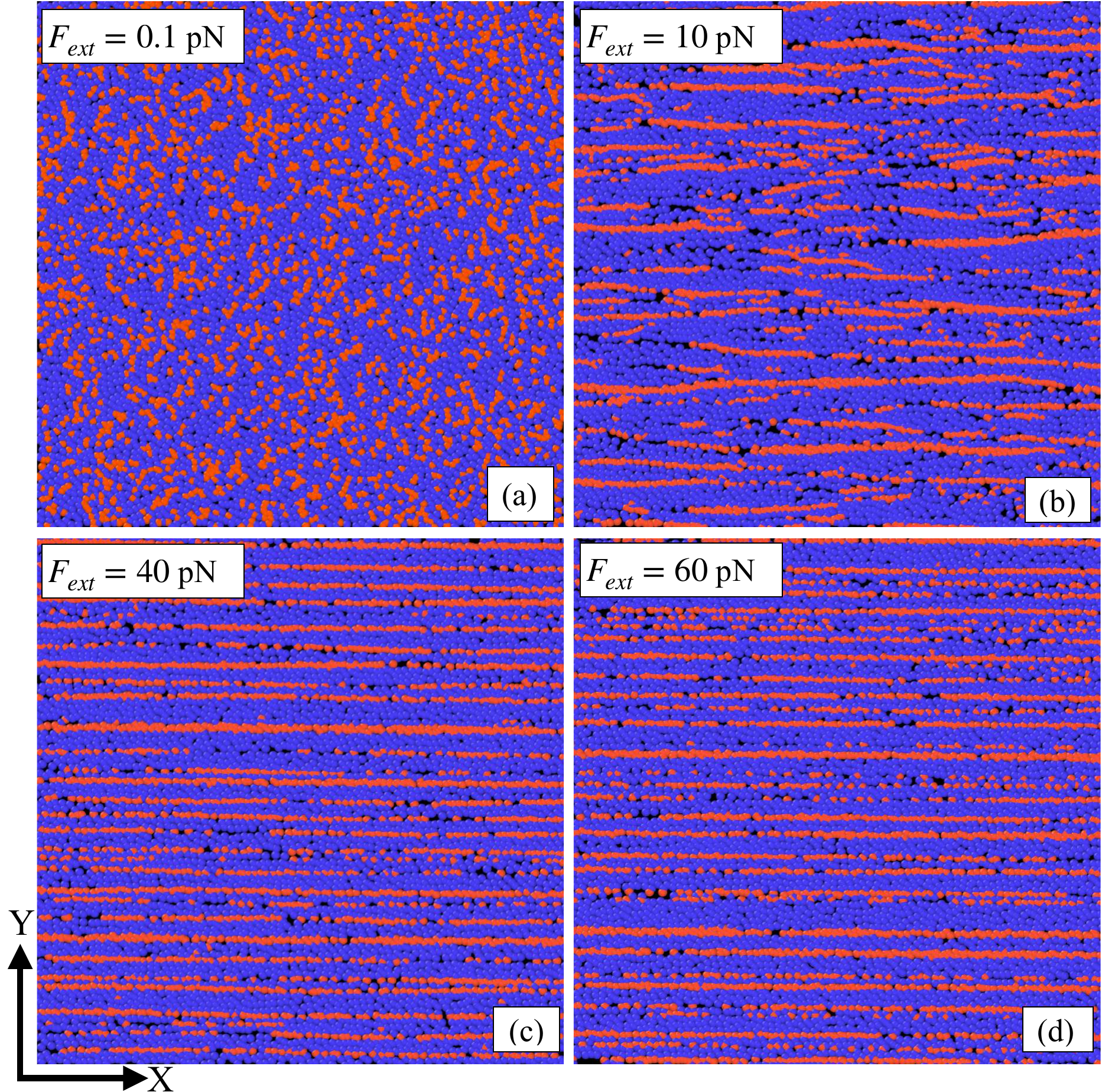}
\caption{The effect of forces on the off-critical mixture with a concentration ratio of 0.3:0.7 at time $t = 50~\omega_{pA}^{-1}$. The simulation carried out for $\Gamma_A =10$ and $\kappa=0.9$.}
\label{fig:fig13}
\end{figure}
\label{sec:app_off_cic}
Here, we present the effect of $F_{ext}$  on segregation in an off-critical mixture with a concentration ratio of $A :B = 0.3:0.7$.
Figure~\ref{fig:fig13} shows snapshots of the binary mixture under different external forces: $F_{ext} = 0.1, 10, 40$,  and  $60$  pN in subplots (a)–(d), respectively.
At very low forcing ($F_{ext} = 0.1$ pN), segregation is insignificant. However, as $F_{ext}$ increases, segregation becomes noticeable, similar to the case of a critical mixture. Due to the lower concentration of species A, laning occurs earlier than in the critical mixture.
From the variation of  $v_d$  as a function of $\Psi_A$ discussed in Sec.~\ref{subsec:drift_gamma_phi}, we conclude that at this concentration ratio of 0.3:0.7, the segregation process does not differ significantly from that of the critical mixture.
\section*{Appendix B: Lane/band formation in 3-dimensional binary critical mixture}
\label{sec:app_3D}
\begin{figure}[t]
\includegraphics[width=\linewidth]{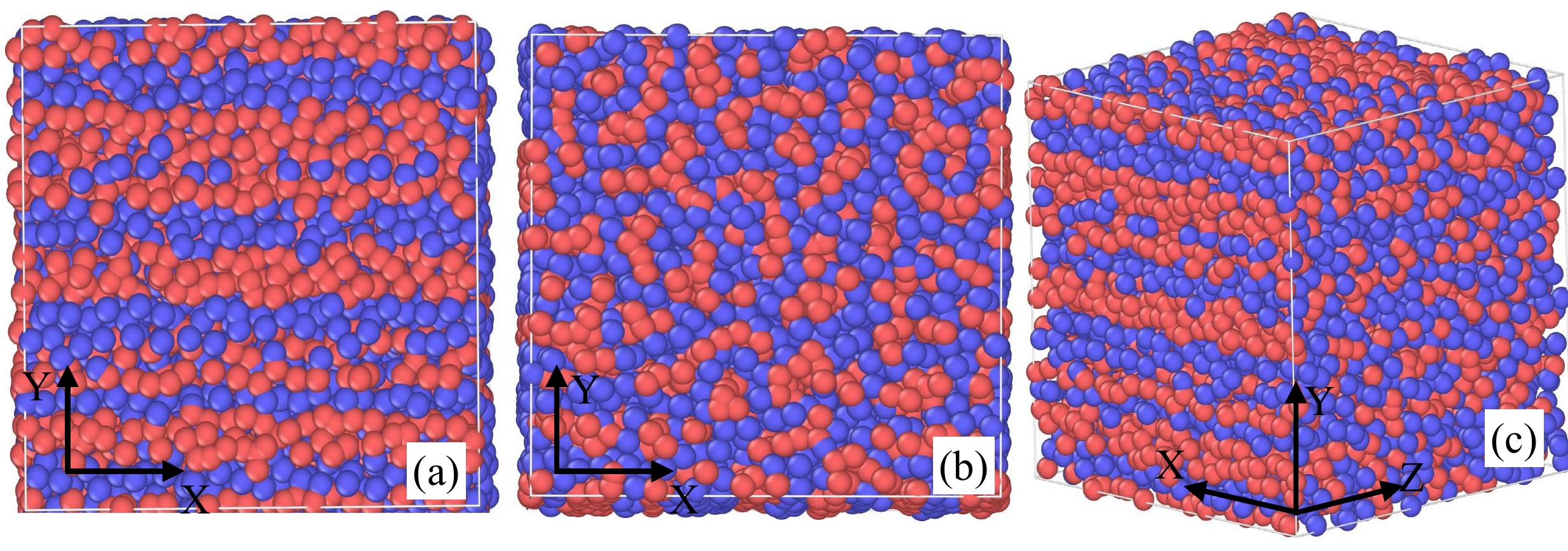}
\caption{Segregation of same-type species ( as in the 2D case) in a 3D binary dusty plasma mixture at $t= 170~\omega_{pA}^{-1}$ for $F_{ext} = 80$ pN. Subplots (a), (b), and (c) show the segregation in XY, YZ, and XYZ orientations, respectively.}
\label{fig:fig14}
\end{figure}

We have also performed a three-dimensional (3D) simulation of the binary dusty plasma and observed segregation in the system.  Figure~\ref{fig:fig14} provides segregation in different planes. Like in the 2D simulation, we apply external force only in the X-direction on type A particles. The subplot (a) shows that the lane is forming in the XY plane. 
Subplot (b) shows that there is no lane in the YZ plane. Moreover, subplot (c) shows all three planes of the cubic simulation box.
%
\end{document}